# Experimental investigation of alkaline treatment processes (NaOH, KOH and ash) on tensile strength of the bamboo fiber bundle

Jerachard Kaima[1, 2, *], Itthichai Preechawuttipong[1], Robert Peyroux[3], Pawarut Jongchansitto[1], Tawanpon Kaima[1]

[1]Department of Mechanical Engineering, Faculty of Engineering,
Chiang Mai University, 50200 Chiang Mai, Thailand
[2]Ph.D. Program in Mechanical Engineering, Faculty of Engineering, Chiang Mai University,
50200 Chiang Mai, Thailand
[3] Univ. Grenoble Alpes, CNRS, Grenoble INP, 3SR, 38000 Grenoble, France
* Corresponding Author: jerachard_k@cmu.ac.th

## Abstract

The aim of this experimental research is to investigate the effect of different alkali solutions, their different concentration, and the soaking time on tensile strength of bamboo fiber bundles. The results of this research will suggest the method to extract bamboo fiber in order to develop biodegradable composite materials in the future. Bamboo strips were mechanically separated from Rough Giant Bamboos, and they were soaked in three different alkali solutions based on NaOH, KOH, and ash with three individual concentrations (5%, 10%, 20% w/w) for eleven different soaking times. The bamboo strips were then washed and naturally air-dried before separating bamboo fibers bundles (cross-section area between 0.06-0.27mm$^2$) by hand pull off. The tensile strengths of bamboo fiber bundles were determined according to ASTM D3039 standard whereas paper grips technique was applied to prevent damage from the machine clamps. The results suggest that the optimal time for soaking bamboo strips is 48 hours for all solution and concentration. Moreover, soaking bamboo strips in KOH, and NaOH with more than 5% (10%, 20%) concentration, bamboo fiber bundles were destroyed. On the other hand, soaking bamboo strips in all concentrations of ash solution, and 5% concentration of KOH, and NaOH provides the same tensile strength value as the one with the soaking bamboo strips in distilled water. Additionally, the alkali solutions have proven to be able to protect bamboo fibers from fungus growth in contrast to distilled water.

**Key word** Bamboo fiber bundle; Chemical extraction; Bio fiber; Strength of fiber; Paper grips

## Introduction

Asian people have extensively learnt how to use bamboo in a variety of applications [1-3]. For instance, kitchenware such as chopsticks, dishes, cups, etc, were largely made from bamboo. In terms of transportations, bamboo was applied as a framework for concrete road, or in the design of rafts, and rowboats. Moreover, bamboo was used in building structures such as anchor piles or trusses [2, 3]. A main reason why bamboo has been selected in many applications is its great performance in mechanical properties [4-8], its variety of species [2, 4, 9], and its rapid growth [6, 10-12]. However, the drawback of the bamboo is its short lifetime in terms of usability (4-5 years) [13]. In addition, the bamboo products are sensitive to insect attacks and degradation by fungi [13, 14]. Even if these enemies do not attack the bamboo fibres directly, they destroy the starch that binds the fibres together and therefore alter the mechanical strength of the whole. In order to solve these problems, mechanical and chemical methods were applied [3]. It is also worth noting that today there is a growing interest in extracting bamboo fibres for use as reinforcement in composites [8, 15].

Several mechanical methods have been applied for cutting the bamboo into small strips. Slicing is one of the most popular methods, which provides a rather constant cross-section [16-18]. However, this method is only appropriate for handmade production of small series because high cutting skill is required. Another widely used mechanical method is the stream explosion. In this method, bamboos are exploded into bamboo strips in a reactor. A collection of bamboo strips with various cross-sectional shapes are obtained [19-21]. Nevertheless, the drawback of the mechanical methods is the size of the bamboo strips obtained which remains excessively large and cannot be employed as a reinforcement of composite materials [22-24]. Therefore, many chemical methods have been employed to solve this the problem [25]. One of the most common methods used is the soaking of bamboo strips in a sodium hydroxide (NaOH) solution for separating the bamboo fibers [10, 26-28]. Note that the NaOH solution was extensively used in many industries, due to its low cost and because a specific skill not required. Nevertheless, the use of the NaOH solution can lead to overreaction, and not only starch but also fiber strength can be altered [29].

Based on a local and traditional knowledge existing in the north region of Thailand, the starch can be eliminated from the wood by using a solution prepared with wood ash. The wood ash can be obtained in a clay stove or a fireplace, which is generally used for local home cooking. In this case, environment effects such as temperature, humidity, etc. cannot be completely controlled, thus leading to its chemical structure differing from the other solutions used in the chemical methods. Moreover, the use of the ash solution for eliminating the starch has been rarely found in the literature. Therefore, the objective of this study is to investigate the effect of alkaline solutions prepared by NaOH, KOH and ash with three different concentrations (5%, 10%, and 20% w/w) on the tensile strength and the optimal soaking time. The paper is organized as follows. Section 2 describes a specimen preparation and an experimental setup. Section 3 is devoted to the analysis and discussion of the results obtained in this study.

**Experimental Setup**
2.1 Preparation of a bamboo fiber bundle

Rough Giant Bamboo (*Dendrocalamus elegans*) older than two years old [30] and have grown in Mae Tang, Chiang Mai, Thailand were used in this experiment. The bamboo stem can be separated into three parts: top, middle, and bottom parts [17, 18, 31]. Only the middle part of the bamboo (1.5 - 3 m from the ground) was cut into several internodes for the sample preparation used in this study, as shown in figure 1a [3, 17, 18, 21, 25, 32], due to the smaller presence of joints and the higher concentration of strong fibers compared to the other parts [16, 25, 32]. Each internode of the cut bamboo was then dried at the outdoor ambient with temperature 36 ±1ºC for 7 days, as shown in figure 1b [26]. The dried internodes were carefully sliced along the longitudinal axis (see in figure 1c), in order to obtain a bamboo strip with a given size. Figure 1d presents the dimensions of the bamboo strip, which is $5 \times 1 \times 60$ mm (thickness×width×length). The strips were then cleaned by a distilled water and finally dried at room ambient temperature [17].

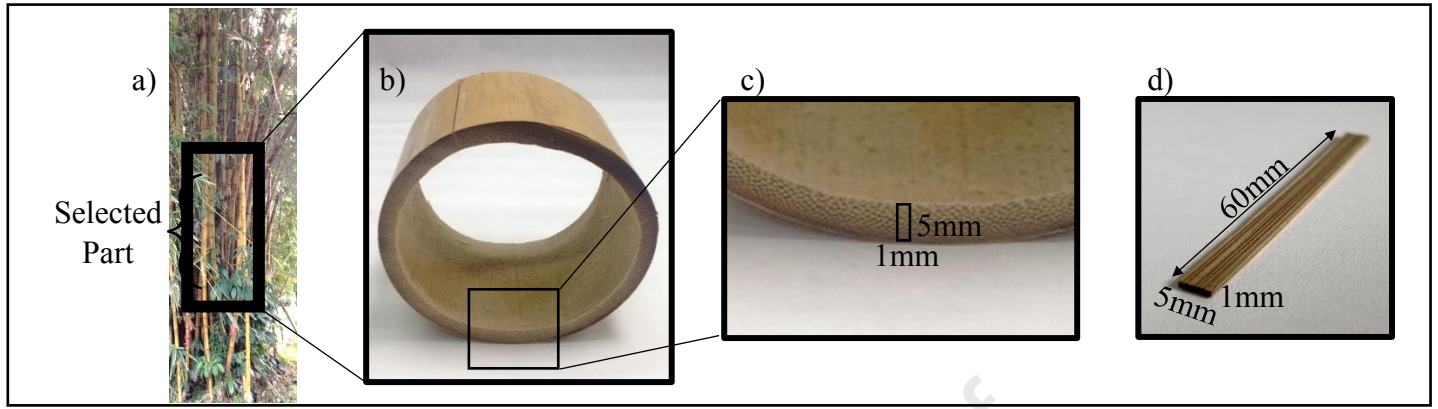

**Figure 1.** Preparation of bamboo strip: (a) the middle part of bamboo, (b) dried internodes of bamboo. (c) cross-sectional area of bamboo strip, and (d) dimensions of bamboo strip.

In order to use natural fibers as reinforcement in a composite material, they have to be separated in individual fibers or at least in bundles of fibres [31]. The most popular method for separating natural fiber is the chemical treatment [33]. Three different alkali solutions prepared respectively from NaOH, KOH and ash, were employed for soaking the bamboo strips. NaOH and KOH solutions are normally used for extracting natural fiber in industrial processes [22, 33] while ash solution is a traditional solution used to extract fiber in north region of Thailand. The ash used in this study was obtained from longan wood (as shown in figure 2) to be as closed as possible to the traditional method. These chemical substances were in a solid state, and in order to prepare the alkali solutions at different concentrations, they were mixed with distilled water. The concentration of the solution can be determined by

$$\%Sol = \left(\frac{(M_{Sub} \times \%Pure)}{(M_{Sub} \times \%Pure) + M_{WATER}}\right) \times 100 \qquad (1)$$

where *%Sol* is the concentration of the solution, $M_{Sub}$ is the mass of the substance (g) *and* $M_{WATER}$ is the mass of the distilled water (g), and *%Pure* is the purity of the substance.

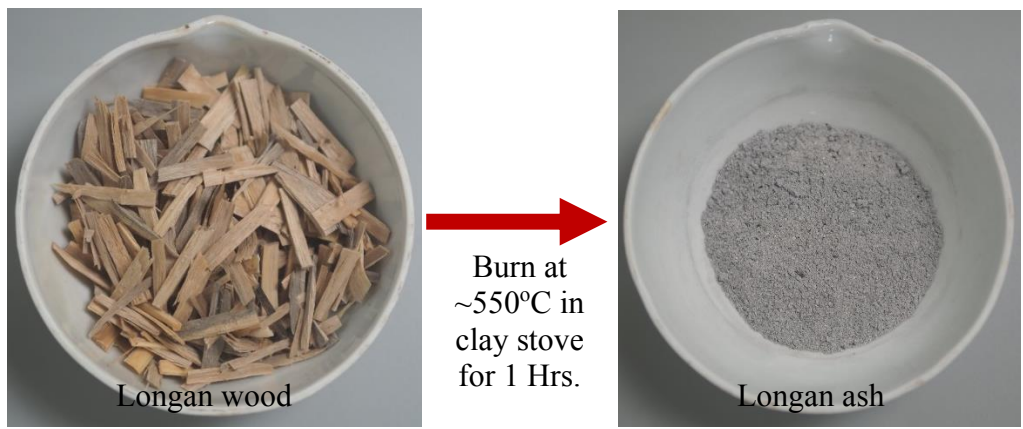

**Figure 2.** Longan ash preparation

For the preparation of the alkali solutions, these substances were slowly and constantly added to avoid a heavy chemical reaction. After that, the alkali solutions were stirred, and they were leaved for 1 day for sedimentation. Only clear solutions were considered in this experiment. Table 1 presents pH values of each alkali solutions used in this experiment where the concentration of each alkali solutions was deliberately varied (5%, 10%, and 20%), in accordance with the traditional concentration of ash solution used in north region of Thailand and with concentrations of NaOH solutions recommended in the previous studies [21, 26, 28].

**Table 1.** Purity of substance for making solution and pH value of each solution.

| Substance | Purity of substance | Concentration (%) | pH Value |
|---|---|---|---|
| Ash | 100% | 5 | 12.2 |
| | | 10 | 12.5 |
| | | 20 | 12.7 |
| KOH | 87.5% | 5 | 13.3 |
| | | 10 | 13.5 |
| | | 20 | 13.7 |
| NaOH | 98% | 5 | 13.1 |
| | | 10 | 13.2 |
| | | 20 | 13.5 |
| Distilled Water | 100% | 0 | 7.0 |

The procedure to obtain bamboo fiber bundles from bamboo strips can be summarized in a flowchart shown in figure 3. Five bamboo strips were soaked in each small close bottles for one of the concentrations of each alkali solution with the liquor-to-wood ratio of 10:1 (w/w). The samples were then let under room ambient temperature (25°C) until the end of the soaking time. It is interesting to mention that the effect of the soaking time on the tensile strength of the fiber has never been reported in the previous studies. Therefore, in order to investigate that effect on the bamboo fiber bundle, the soaking time used in this study for each solution was chosen as presented in Table 2. It can be seen that soaking times less than 24 hours in distilled water were not considered. This is due to a recommendation proposed by Garcia, *et al.*[34] saying that the soaking time in distilled water for the bamboo should be more than 3 days. The soaked samples were then cleaned in distilled water, in order to stop the chemical reaction [11, 26, 28]. Afterward, the samples were dried under the ambient temperature (25°C) for 7 days before packing them in the plastic bags for preventing the humidity effects [26]. Before performing mechanical tests, the bamboo fiber bundles were obtained by manually tearing off from each sample. The cross-sectional area of the fiber bundles was measured in a range from 0.06 mm$^2$ to 0.27 mm$^2$ with the length of 60 mm [17, 35]. This manually tearing off was also done in another research [13, 17].

**Table 2. Soaking time used for each solution.**

| Solution type | Soaking Time (hrs.) | | | | | | | | | | |
|---|---|---|---|---|---|---|---|---|---|---|---|
| | 1 | 3 | 6 | 9 | 12 | 18 | 24 | 48 | 72 | 120 | 168 |
| Ash | ✓ | ✓ | ✓ | ✓ | ✓ | ✓ | ✓ | ✓ | ✓ | ✓ | ✓ |
| KOH | ✓ | ✓ | ✓ | ✓ | ✓ | ✓ | ✓ | ✓ | ✓ | ✓ | ✓ |
| NaOH | ✓ | ✓ | ✓ | ✓ | ✓ | ✓ | ✓ | ✓ | ✓ | ✓ | ✓ |
| Distilled Water | ✗ | ✗ | ✗ | ✗ | ✗ | ✗ | ✓ | ✓ | ✓ | ✓ | ✓ |

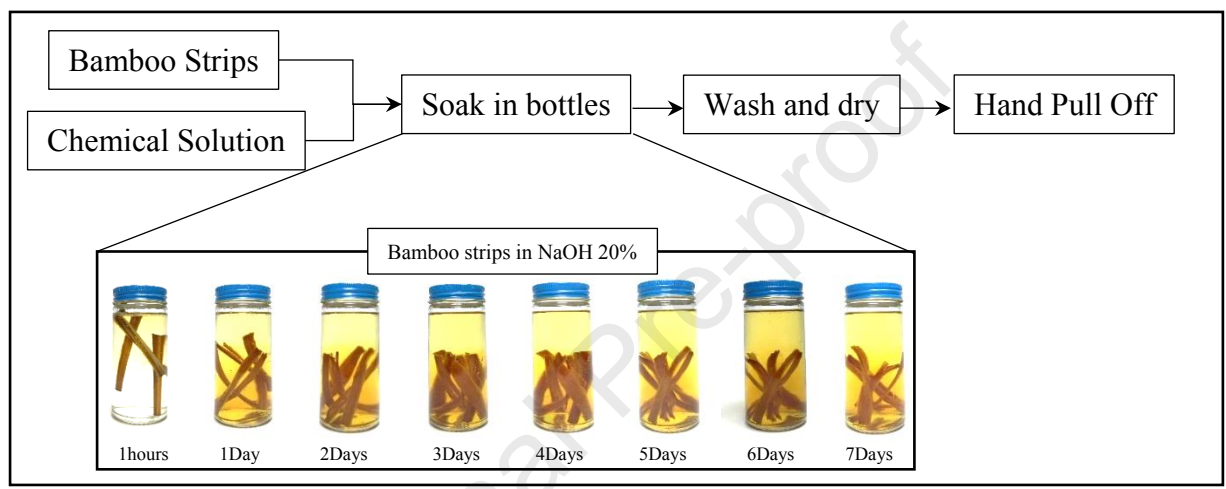

**Figure 3.** Preparation scheme of bamboo fiber bundle from bamboo strip.

2.2 Tensile tests of the bamboo fiber bundle

In this section, a procedure to determine the tensile strength of the bamboo fiber bundle is described according to ASTM D3039 standard. In order to avoid a damage of the fiber bundle at the clamping point of the testing machine, a paper grip technique was applied for this purpose [19-21, 36, 37]. The paper grip was made from an 80 grams paper, which was cut following figure 4a. The yellow area shown in figure 4b was an area for gluing. The fiber bundle represented by a green solid line was then placed along the symmetric vertical axis of the paper grips. Since the free length of the fiber bundle must be equal to 3 cm, the fiber bundle was also symmetrical about the cutting line. Afterwards, parts A and D of the paper grips were flipped over black dashed lines. To ensure that the glue was completely dried, the paper grip was left under the ambient temperature for a day. Finally, the fiber bundle attached with the complete paper grip, which is here referred to as a specimen, was achieved.

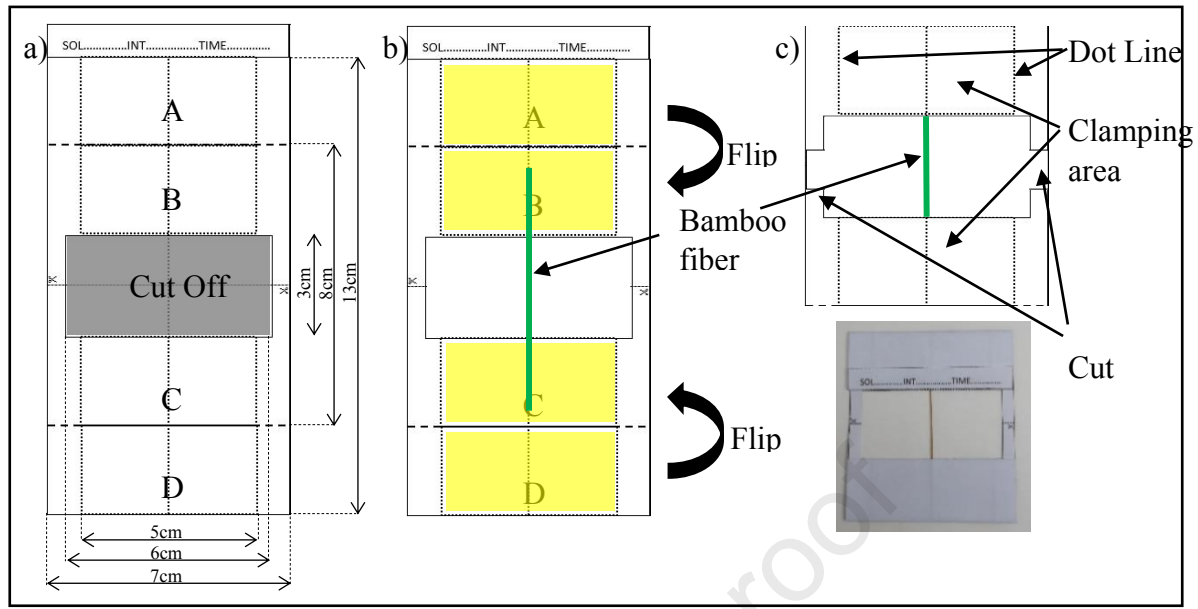

**Figure 4.** Paper grip technique: (a) Layout of paper grip, (b) attaching the fiber on paper grip, and (c) complete paper grip with the fiber.

A clamping area of the specimen (see in figure 4c) was carefully clamped in INSTRON 5566 testing machine. Before starting the experiments, it is necessary to cut the remaining paper within the free length zone along a black dotted line with the scissors symbol as shown in figure 5. A uniaxial tensile loading with a displacement-controlled of 1 mm/s was then applied until the tear point of specimens. Only the fiber bundle with its tear points over 25% of the free length from the clamping area was considered, as shown in figure 6a and 6b. Examples of unconsidered specimens are presented in figures 6c and 6d. The tensile strength of those considered specimens was then determined. In addition, more than 5 specimens of each case were employed to average the tensile strength of the fiber bundle.

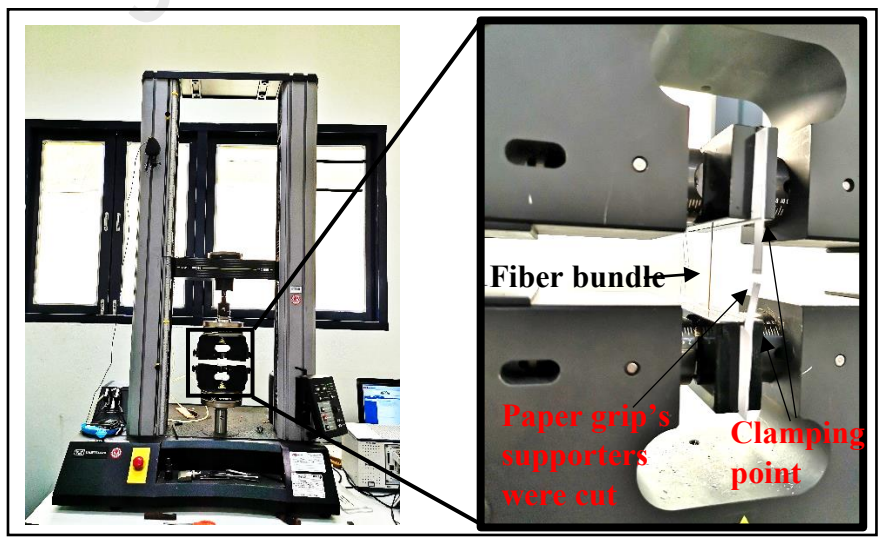

**Figure 5.** Experimental setup.

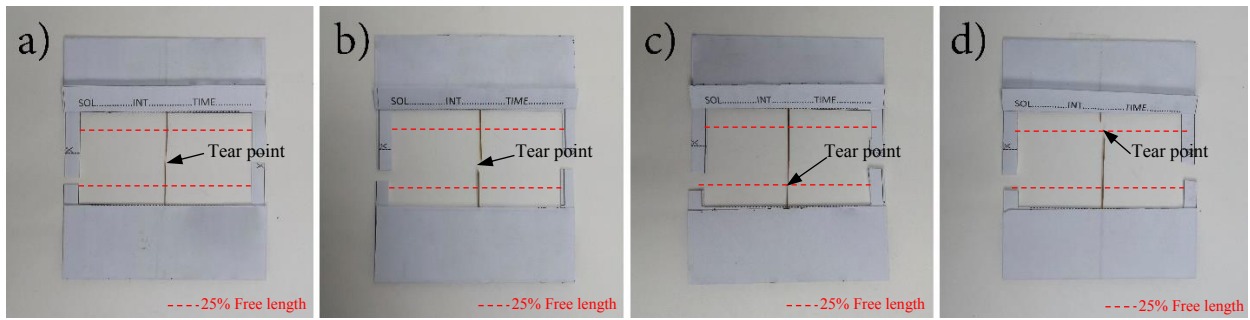

**Figure 6**. Examples of selected case and non-selected case
a) considered case b) considered case c) unconsidered case d) unconsidered case

## Experimental results

The tensile strength values of the bamboo fiber bundles are described as the function of soaking time, alkali type and the concentration.

*Effect of soaking time on tensile strength of bamboo fiber.*

Figure 7 shows the result of bamboo fiber bundles tensile strength in each soaking time. The circle, triangle, square, and diamond represented NaOH, KOH, ash solution, and distilled water respectively. The tendencies of tensile strength are represented by dot line, dot-dash line, and dash line for NaOH, KOH, and ash solution respectively. The colors of each trendline and sign stand for the concentration of the solutions (blue, red, and black for 5%, 10%, and 20% respectively). The tensile strength of bamboo fiber bundle in each treatment converges to distinct values after 48 hours soaked as show in table 3. The alkali solutions gradually infiltrate into bamboo strips after 48 hours soaked, so the outer and inner bundles surfaces were completely moist and homogeneous. The tensile strength in each treatment distributed before 48 hours as shown in figure 7. So, the tensile strength values of bamboo fiber that soaked less than 48 hours are not considered. As a result, 48 hours is the sufficient time for soaking bamboo strips in alkaline solution. The average tensile strength values of bamboo fiber bundle that show in this report were calculated by using the tensile strength value from bamboo fiber bundles soaked over 48 hours only.

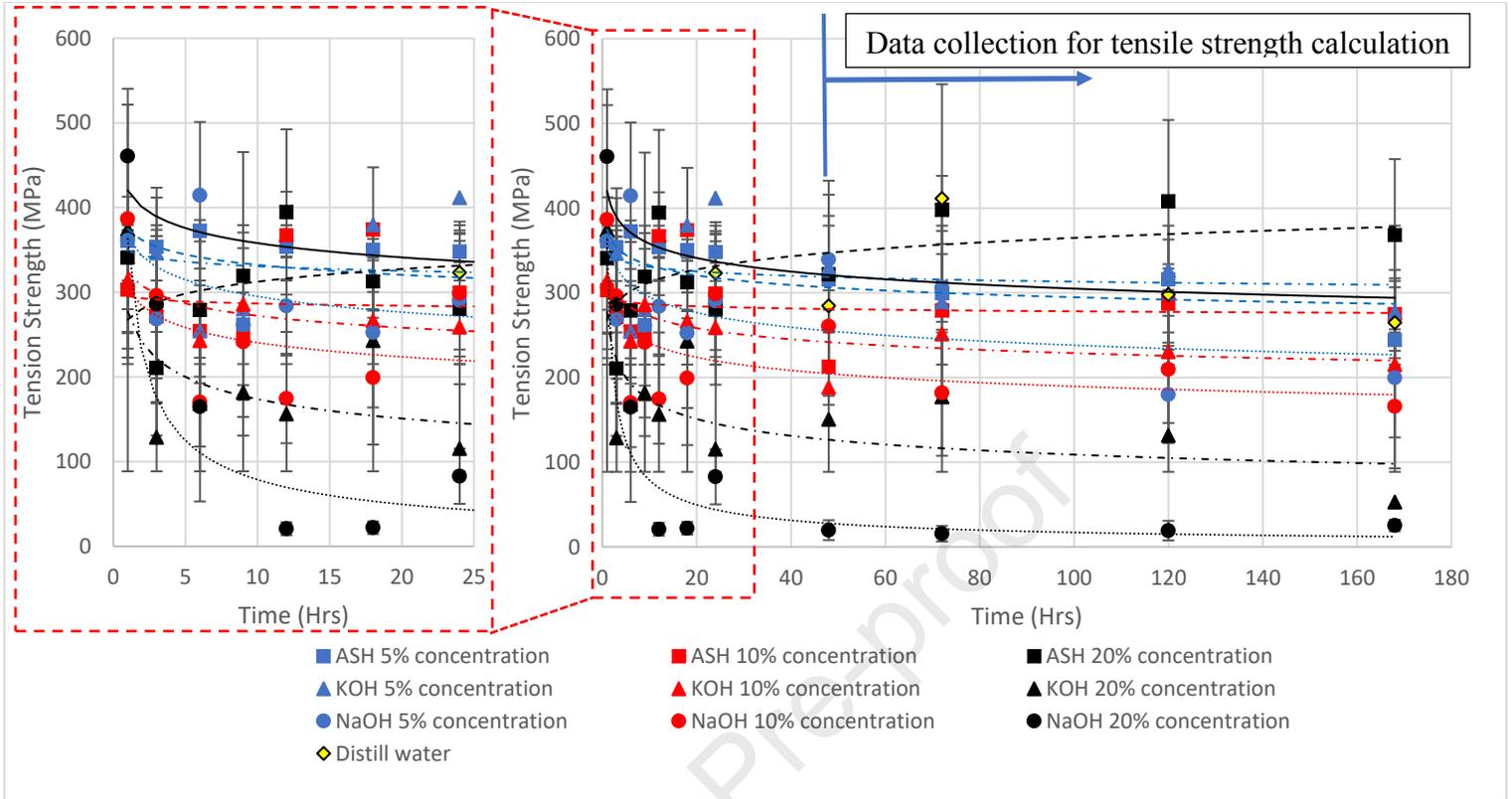

**Figure 7.** Tensile strength of bamboo fiber bundles VS soaking time with different alkaline solution and concentration.

**Table 3. Average tensile strength of bamboo fiber value after 48 hours soaking time in individual cases.**

| Solution | Concentration (%) | Tensile strength (MPa) |
| --- | --- | --- |
| Ash | 5 | 293±17 |
|  | 10 | 263±17 |
|  | 20 | 374±19 |
| KOH | 5 | 305±13 |
|  | 10 | 221±13 |
|  | 20 | 128±26 |
| NaOH | 5 | 256±39 |
|  | 10 | 204±20 |
|  | 20 | 19±2 |
| Distilled Water | 0 | 314±32 |

*Effect of alkaline type on tensile strength of bamboo fiber bundle.*

This part shows the effect of alkaline solution type on the tensile strength of bamboo fiber bundle. The tensile strength values were averaged after 48 hours, and we only present the 20% concentration in the individual type of solution as show in figure 8. The tensile strength of bamboo fiber bundles soaked in distilled water, which is 314±32 MPa, is used to compare to the others. The tensile strength fiber bundles soaked in distilled water are found to be similar as the ones in the previous report from Lau [7].

However, fungus growth on the bamboo strips skin, as shown in figure 9a, appeared while soaking in distilled water before naturally air-drying and keeping in the plastic bag over 7 days [3, 34, 38]. This fungus destroyed the skin of the bamboo fiber bundle and disrupted the hand pull off method. As a consequence, the bundle fibers teared at the loci where the fungus grown. Indeed, Garcia [34] reported the same result while soaking the bamboo in water and suggested that using running water should solve fungus growth problems.

Moreover, the approximate tensile strength result was provided by the ash solution. The tensile strength of the bamboo fiber bundles soaked in the ash solution is 374±19MPa. Only middle lamella (pectin, wax, calcium, etc.) in bamboo fiber strips was eliminated, so the processed bamboo fiber bundles direction still arranges in the longitudinal as show in figure 9b. This is benefit for the hand pull off method before testing the fibers bundle. The fiber bundles cross-sectional area was constant throughout the bundle fiber length. Note that there is no fungus growth on the bamboo skin. The tensile strength of the bamboo fiber bundles that soaked in the KOH solution and the NaOH solution are found to be less than the others. The average tensile strength of bamboo fiber bundles values for the KOH and the NaOH are 128±26 MPa and 19±2 MPa, respectively. Not only removed middle lamella but also the fiber surface was heavy destroyed. The results are in agreement with Ferreira, Cao, Jahan and Puls [10, 39-41]. Moreover, bamboo fibers were twisted by chemical reaction as shown in figure 9c-9d. This phenomenon affects to the shape of bamboo fibers bundle. They did not present a continuous cross-section, and they were too short for the tensile strength test. These also affects to the difficult tearing by hand pull off step.

Hence, it can be concluded that the type of solution may not directly affect to the tensile strength, but the concentrations of the solution may have a significant effect on the tensile strength (will be discussed more detail in the next section). However, the chemical solution types contribute to reduce the soaking time, to prevent from fungus growth, and make it easier to extract the bamboo bundles from the stipes.

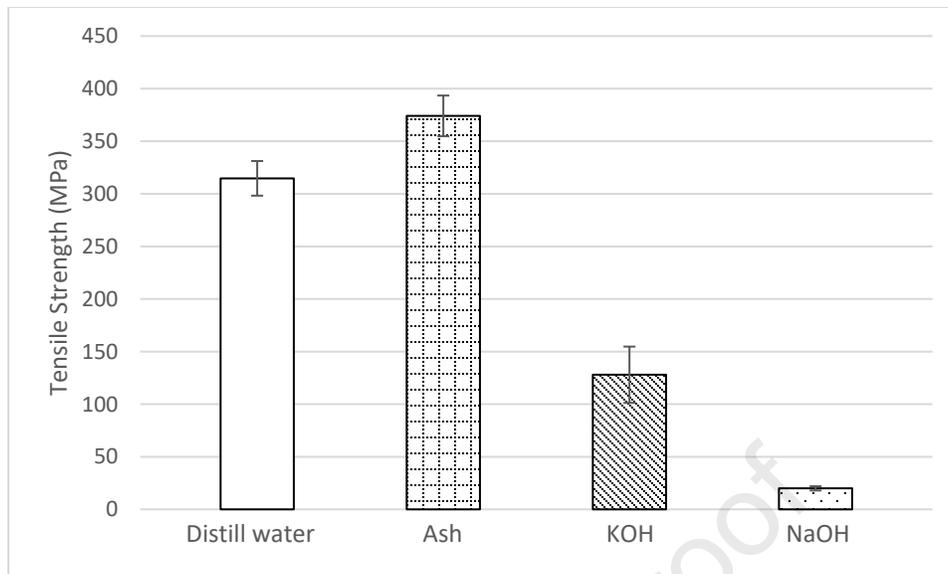

**Figure 8.** The average tensile strength of bamboo fiber that soaked in 20% concentration with individual alkaline type after 48 hours

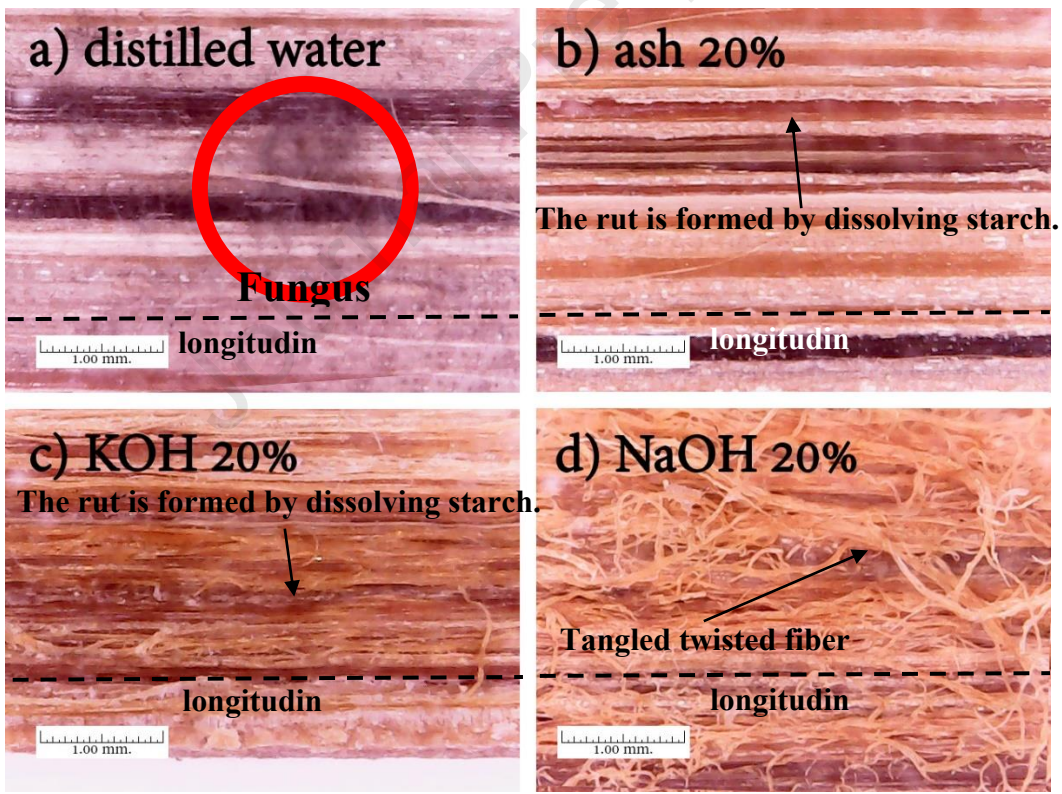

**Figure 9.** Surface of bamboo fibers soaked in difference type of alkali solution at 20% concentration for 48 hours. a) distilled water b) ash solution c) KOH solution d) NaOH solution

*Effect of concentration in different alkali solution on tensile strength of bamboo fiber.*

Figure 10 shows the average tensile strength of bamboo fiber bundles that were soaked in individual concentration of different alkali solution after 48 hours. The tensile strength of bamboo fiber bundles soaked in distilled water is 314±32 MPa was used as a control sample (The value is in the same range with Lau [7]). For all types of alkali at 5% concentration, the tensile strength of bamboo fiber bundle is not affected. For all concentration of ash solution, tensile strengths of bamboo fiber bundles were close to a control sample value. The surface of bamboo strips soaked in 5% and 10% ash solutions are similar to surface of a control sample. The fibers are still well arranged in the longitudinal direction, as shown in figure 9a and figure11a-b. However, in the case of soaked bamboo strips in 20% ash solution, there is more rut on the bamboo strip as show in figure 11c which the middle lamella is removed. The ruts on bamboo strip offered good result in equably hand pull off that provide the constant cross-section of fibers bundle.

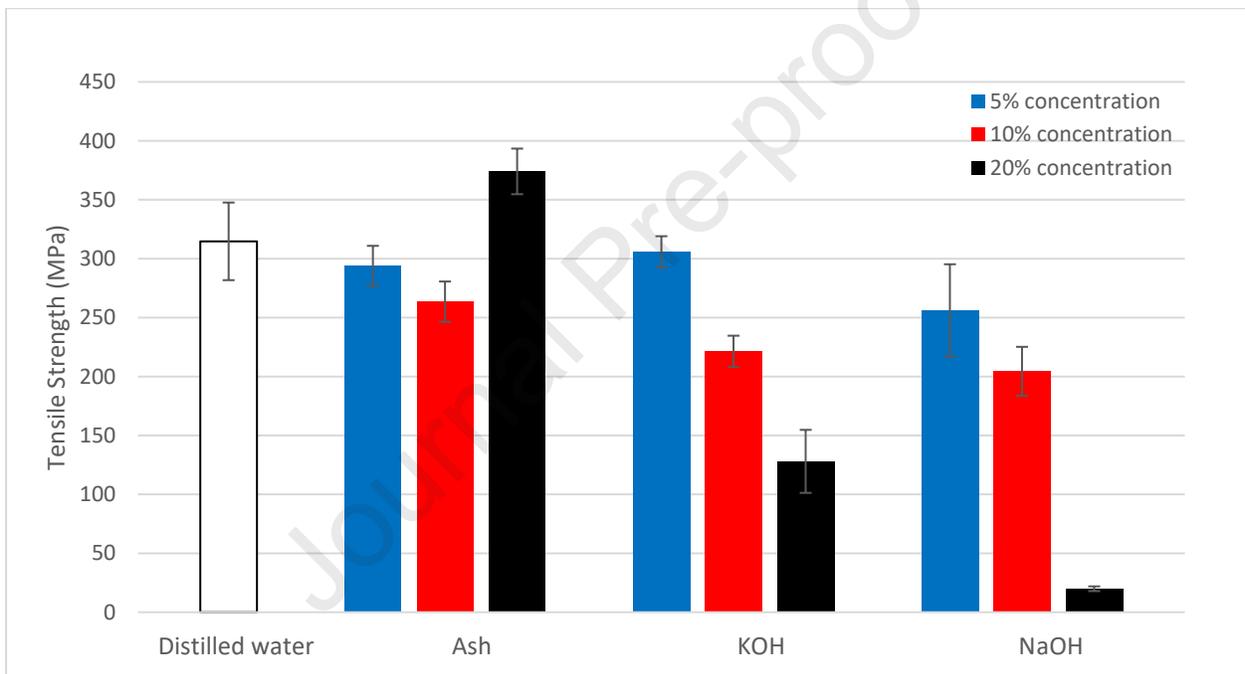

**Figure 10.** The average tensile strength of bamboo fiber after 48 hours soaked in individual alkaline type, and different concentrations.

Furthermore, the tensile strength of bamboo fibers bundle that soaked in the KOH and the NaOH solution were significantly reduced while the concentrations were increased [42]. The surface of bamboo strips that soaked in the KOH solution, and the NaOH solution are shown in figure 11d-i. In case of soaking the bamboo strips in the 5% KOH solution, or in the 5% NaOH solution, only middle lamella was removed, and the fibers bundle were still oriented in the longitudinal direction as shown in figure 11d and figure 11g. This phenomenon is similar to soaking the bamboo strip in the distilled water as shown in figure 9a. The result is in agreement with Zhang, Olakanmi and Ku [26, 27, 42]. Focusing on the 10% KOH and the 10% NaOH solutions, some fibers were slightly extracted from the bamboo strip as shown in figure 11e and figure 11h. In the case of 20% KOH solution, some of them were teared and were more unidirectional as shown in figure 11f. In the same way, the bamboo strips surface was extremely altered and ruined by the 20% NaOH solution as shown in figure 11i. This result is in agreement with Sanchez [28], The hand pulls out method was directly affected by the shape of fibers bundle because only the long straight fibers bundle was easily pulled out from the unidirectional fibers strip. Those fibers bundle must be long enough to be tested in the tensile test machine. However, the puffy fibers bundle was too short, and they were destroyed while extracting by the hand pull off method.

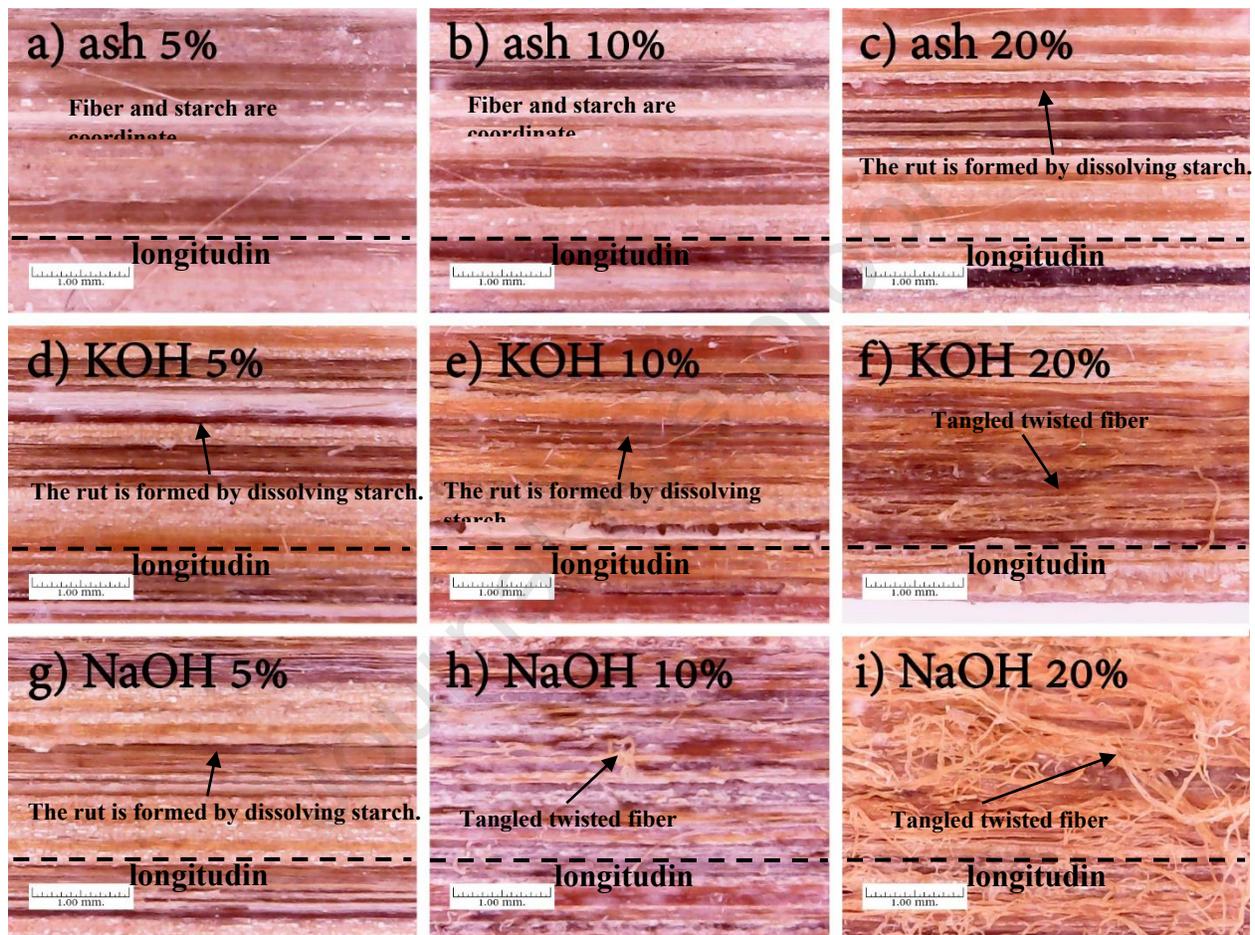

**Figure 11.** Surface of bamboo strips soaked in different concentration and alkali solution

## Conclusion

In order to investigate the optimal soaking time, the suitable chemical solution, and the concentration values for extracting the bamboo bundle as reinforcement in composite, the NaOH solution, the KOH solution, and the ash solution with the concentration 5%, 10%, and 20% w/w were used to soak bamboo strips for defining the best treatment. The bamboo strips size 1mm x 5mm x 60mm were soaked in these solution for 1 hours to 7 days. The bamboo strips were then separated by the hand pull off method for producing the bamboo fiber bundles. The tensile strength of bamboo fiber bundles was tested following the ASTM D 3039 with the paper grips technique to avoid the damaging of the fibers bundle at the gripping points. The tensile strength was described under the influence of the soaking time, type of solution, and the concentration of the solution

- 48 hours is the optimal soaking time for all alkali solutions since the bamboo strips are absolutely permeated by the solution. Moreover, the tensile strength of bamboo bundle has not been disrupted from the alkali solution.

- For the KOH and the NaOH solutions, the tensile strength of bamboo fiber bundle decreases dramatically when the solution's concentration is more than 5%, due to bamboo fiber destruction, the discontinuity and the misalignment of fibers.

- 5% concentration of all alkali solutions (ash, KOH, NaOH) do not affect the tensile strength of bamboo fiber bundles comparing with soaking in distilled water.

- For the natural ash solution, the solution's concentration does not affect the tensile strength of bamboo fiber bundle. Moreover, soaking bamboo in the ash solution provides the benefit on the fungus growth protection and the constancy cross-section area of bamboo fiber bundle.

Therefore, this experiment suggests that all concentration of the ash solution and the 5% concentration of the NaOH and the KHO solutions must be used as the alkali solution for extracting the bamboo fiber bundles for fabricating composite.

## Acknowledgments

The authors gratefully acknowledge Research and Researchers for Industries Scholarship (Grant No. PHD 60I0039) and Charoen Triphop Limited Partnership for their financial support during the research. The authors also thank Mr. Pichan Powpongchan for his suggestions about the chemical mixing technique.

**Highlights**
- Bamboo fiber extraction treatment by soaking in alkali solution were scientifically investigated.
- The suitable time for soaking bamboo strips in alkali solution was reported.
- Ash solution, which is the local and non-toxic substance, was represented as the treatment to provide the natural fiber with 100% eco friendly during the manufacturing process.
- Using alkali solution not only encourage fiber extraction but also prevent fungus growth on the bamboo strip skin.

**Declaration of interests**

☐ The authors declare that they have no known competing financial interests or personal relationships that could have appeared to influence the work reported in this paper.

☒ The authors declare the following financial interests/personal relationships which may be considered as potential competing interests:

Jerachard Kaima reports financial support was provided by Research and Researchers for Industries Scholarship, Thailand.